\documentclass[twocolumn,preprintnumbers,amsmath,amssymb]{revtex4}

\usepackage{graphicx}
\usepackage{dcolumn}
\usepackage{bm}
\usepackage{xcolor}

\usepackage[a4paper, total={7in, 10.in}]{geometry}

\usepackage{mathtools}

\begin{document}


\title{Effects of the QCD Equation of State and Lepton Asymmetry on
  Primordial Gravitational Waves}

\author{Fazlollah Hajkarim}
\email{hajkarim@th.physik.uni-frankfurt.de}

\author{J\"urgen Schaffner-Bielich}
\email{schaffner@astro.uni-frankfurt.de}

\author{Stephan Wystub}
\email{wystub@astro.uni-frankfurt.de}
  
\affiliation{ Institut f\"ur Theoretische Physik, Goethe Universit\"at, Max von
  Laue Strasse 1, D-60438 Frankfurt am Main, Germany }

\author{Mandy M. Wygas}
\email{m.wygas@physik.uni-bielefeld.de}

\affiliation{Fakult\"at f\"ur Physik, Universit\"at Bielefeld, Postfach
  100131, 33501 Bielefeld, Germany}

\date{\today}

\begin{abstract}
  Using the quantum chromodynamics (QCD) equation of state (EoS) from lattice
  calculations we investigate effects from QCD on primordial gravitational waves
  (PGWs) produced during the inflationary era. We also consider different
  cases for vanishing and nonvanishing lepton asymmetry where the
  latter one is constrained by cosmic microwave background experiments. Our
  results show that there is up to a few percent deviation in the predicted
  gravitational wave background in the frequency range around the QCD transition
  ($10^{-10}- 10^{-7}$~Hz) for different lattice QCD EoSs, or at larger
  frequencies for nonvanishing lepton asymmetry using perturbative QCD. Future
  gravitational wave experiments with high enough sensitivity in the
  measurement of the amplitude of PGWs like SKA, EPTA, DECIGO and LISA can probe
  these differences and can shed light on the real nature of the cosmic QCD transition
  and the existence of a nonvanishing lepton asymmetry in the early universe.
\end{abstract}

\maketitle


\section{\label{sec:intro}Introduction}

The era of observing the universe beyond electromagnetic waves began by
the first gravitational wave (GW) observation of LIGO produced from two merging
black holes \cite{Abbott:2016blz}. This gives the opportunity to probe
phenomena in astrophysics and cosmology which are impossible or difficult to
be observed by photons. The inflationary scenario has been proposed in
cosmology as a solution to the flatness, the horizon, and magnetic monopole
problems \cite{Linde:1981mu,Guth:1980zm}. It was shown that inflation can also
produce primordial gravitational waves (PGWs) as the tensor perturbation of
the metric of spacetime \cite{Starobinsky:1979ty}. This also opens up a
direct way to check the physics of the early universe before big bang
nucleosynthesis (BBN) which until now has been hidden from our sight except
for its possible effects on the cosmic microwave background
\cite{Seljak:1996gy,Lyth:1996im,Spergel:2006hy,Ade:2015lrj}.
Different phenomena like electroweak transition, QCD transition, phase
transition in the dark sector, early matter domination, etc., can be present
before BBN which can produce extra GWs or affect PGWs
produced by the inflationary scenario
\cite{Schwarz:1997gv,Maggiore:1999vm,Mazumdar:2018dfl}.

The equation of state (EoS) of the standard model (SM) can have different 
impacts on PGWs due to possible features coming from the quark--gluon and the
electroweak transitions
\cite{Schwarz:1997gv,Maggiore:1999vm,Mazumdar:2018dfl}. These effects can be
measured by future GW experiments. There are some ongoing and future space
and earth based GW detectors like DECIGO
\cite{Seto:2001qf,Sato:2017dkf}, LISA \cite{Audley:2017drz}, SKA
\cite{Janssen:2014dka}, and EPTA \cite{Lentati:2015qwp}, which can measure the
possible effects of cosmic (phase) transitions in the visible and dark side of
the universe in the relevant frequency ranges.

The thermal effect of the SM on PGWs appears via the trace anomaly and the
energy and entropy density of radiation in the equation of motion for PGWs
produced from inflation. The trace anomaly of the energy
momentum tensor of the SM shows deviations of the EoS from pure radiation
(with $p=\rho/3$) which are due to quantum effects and the nonrelativistic
behavior of SM particles at temperatures below about one third of their
masses \cite{Schwarz:1997gv,Watanabe:2006qe,Schettler:2010dp,Caprini:2018mtu}.
Effects from the trace anomaly in the SM are most pronounced at the QCD 
transition \cite{Aoki:2006we}.

The QCD transition can affect the cosmology of the early universe in
different aspects like its effect on the relic density of dark matter and the
GW spectrum \cite{Drees:2015exa,
  Schwarz:1997gv,Castorina:2018whj,Anand:2017kar,Schwarz:2003du,Stuke:2011wz,Saikawa:2018rcs,Hindmarsh:2005ix,Borsanyi:2016ksw,Capozziello:2018qjs,Li:2018oqf}.
What we know from lattice QCD calculations at vanishing chemical potentials for
baryon, electric charge, and strangeness number is that the QCD transition is a
smooth crossover \cite{Aoki:2006we}.
This is in contrast to first studies on the QCD phase diagram for the
early universe which adopted a first order or second order phase transition
\cite{Witten:1984rs,Asakawa:1989bq}. However, the effect of considering
nonvanishing chemical potentials can slightly change the strength of the transition
and lead to different EoSs compared to the case of vanishing
chemical potentials \cite{Schwarz:2009ii,Stuke:2011wz}.

\begin{samepage}
In the present study we focus on the imprints of QCD on PGWs for the cases
of vanishing and nonvanishing lepton flavor asymmetry in the early universe.
This paper is organized as follows. In section~\ref{sec:primordialgw} we
outline our formalism to compute the relic PGW spectrum and the relevant
thermodynamic relations. Then we discuss the impact of different QCD EoSs, based on different lattice QCD results, on the PGW spectrum in
section~\ref{sec:qcdeos} paying special attention to effects from the charm
quark contribution. Effects on PGWs from nonvanishing chemical
potentials, in particular from a nonvanishing lepton asymmetry, are discussed
in section~\ref{sec:chempot}. Finally, we conclude in
section~\ref{sec:conclusion}.
\section{\label{sec:primordialgw}Primordial Gravitational Waves from Inflation}

The production of GWs by the inflationary scenario in the
early universe can be considered by doing a perturbative analysis of the
Friedmann equations. In standard cosmology the following metric describes the
evolution of the cosmos assuming vanishing curvature which is a reasonable
assumption for an isotropic and homogeneous universe \cite{Dodelson:2003ft}
and matches with observations \cite{Ade:2015lrj}:
\end{samepage}
\begin{equation}
\label{frw}
ds^2=-dt^2+a(t)^2d{\bf{x}} ^2\, ,
\end{equation}
where the relation between the cosmic time $t$ and the conformal
 time $\eta$ can be defined by $dt=a\,d\eta$.
The tensor perturbation equation in the Fourier space which shows the
evolution of PGWs is given by \cite{Dodelson:2003ft}
\begin{equation}
\label{tensor-pert}
h^{''}(k,\eta)+2\mathcal{H}(\eta) h^{'}(k,\eta)+k^2 h(k,\eta)=0\, ,
\end{equation}
where ${}^{\prime} \coloneqq d/d\eta$. The conformal Hubble rate is denoted by
$\mathcal{H}=a^{\prime}/a=aH$. By using $v(k,\eta)=a (\eta)h(k,\eta)$ one
has
\begin{equation}
\label{pgw-tensor}
v^{''}(k,\eta)+\left(k^2-\frac{a^{''}}{a}\right)v(k,\eta)=0\, ,
\end{equation}
with 
\begin{equation}
\label{fried1}
\frac{a^{''}}{a}=\frac{4\pi G}{3}\left(\rho_{\rm{tot}}-3p_{\rm{tot}}\right)\, ,
\end{equation}
where $8\pi G=1/M_{\rm{Pl}}^2$ and $M_{\rm{Pl}}=1.22\times10^{19}$~GeV.
 The quantity in
the parentheses at the right hand side of eq.~(\ref{fried1}) is called the trace
anomaly (or interaction measure) and can be written as follows
\cite{Cheng:2007jq,Bazavov:2014pvz,Bazavov:2017dus}
\begin{equation}
   \frac{I(T)}{T^4}=\frac{\rho_{\rm{tot}}-3p_{\rm{tot}}}{T^4}=T\frac{\partial}{\partial
     T}\left(\frac{p_{\rm{tot}}}{T^4}\right)_{\mu/T}\, . 
\end{equation}
In eq.~(\ref{fried1}) one should consider the total energy and pressure density
with respect to the scale factor taking into account entropy conservation in the early
universe. The entropy density $s_{\rm{tot}}$ can be derived by using thermodynamic relations
which we show below. For this one needs also the Friedmann equation which reads
\begin{equation}
\label{fried2}
H^2=\frac{8\pi G}{3}\rho_{\rm{tot}}\, .
\end{equation}
At any specific time, $t(\eta)$, during the cosmic evolution super horizon
modes can be defined for $k\eta \ll 1$. When the universe expands and modes
enter the horizon, they are identified as sub horizon modes by $k\eta \gg 1$.
The frequency of each mode $k$ can be written as $f=k/2\pi$. The initial
condition for modes outside the horizon that we used to solve
eq.~(\ref{pgw-tensor}) to compute the GW spectrum are
\cite{Dodelson:2003ft,Schettler:2010dp}
\begin{eqnarray}
\label{initcon}
v(k,\eta_{\rm{ini}})&=&\frac{1}{k^{3/2}} \, , \nonumber \\
v^{\prime}(k,\eta_{\rm{ini}})&=&\frac{v(k,\eta_{\rm{ini}})}{r_{\rm{ini}}}\, , \quad
r_{\rm{ini}}~=~\frac{1}{a(\eta_{\rm{ini}})H(\eta_{\rm{ini}})}\, ,
\end{eqnarray}
where the oscillatory factor of the wave function, $\exp(ik\eta)$, is
neglected, as it only affects the phase of the GW, not the amplitude we are
interested in. In choosing these initial conditions only the $k$ dependence is
important for our purpose.

By assuming entropy conservation during the QCD transition and until today,
one can compute the evolution of $\rho_{\rm{tot}}(a)$ by solving eq.~(\ref{fried2})
backward in time, i.e.\ from today ($a_0=1$) to a chosen scale factor $a$ in
the early universe
\footnote{If one does not solve eq.~(\ref{fried2}) in this way it leads to an
  error in the computed PGW spectrum which can be larger than the effects from
  the QCD transition. Basically, since the energy density of radiation has
  changed for a specific scale factor with respect to today, it leads to a
  shift in the amplitude for a given frequency.}.
Then the solution can be used to solve eq.~(\ref{pgw-tensor}).

Due to the large range of the scale factor from today ($a_0=1$) back to the
early universe when we consider modes well before horizon crossing
(temperatures well above the electroweak transition) numerical problems for
solving the differential equation given by eq.~(\ref{pgw-tensor}) might arise.
Therefore, one solves the differential equation either until horizon crossing
for each mode or until a temperature after neutrino decoupling to include all
the evolution of the EoS in the calculation. In practice both of these ways of
calculating the GW spectrum will give approximately the same final result. Since the slight
change of the EoS due to the change of trace anomaly
happens in a short interval of the scale factor with a tiny deviation from the
radiation-like EoS, this will cause a tiny change in the amplitude of
the GW when the mode enters the horizon until the end of
neutrino decoupling. This procedure is sufficient for our goal to show the
effects from QCD and from a lepton asymmetry on PGWs. 
We also check the difference between two procedures in 
a specific frequency ($3\times10^{-11}$ and $5\times10^{-8}$ Hz) range such that we can find the numerical solution in a precise way. Definitely, if one finds
any evidence of PGWs in experiments a more detailed calculation can be done by
fixing the scale of inflation from the data to have a more precise handle on
the aforementioned effects to match theory with experiment.
 
For each polarization mode ($\lambda$) of the GW
eq.~(\ref{tensor-pert}) is valid and the amplitude of perturbations can be
written as \cite{Caprini:2018mtu}
\begin{equation}
\label{ampvh}
h_{\lambda}(k,\eta)=h_{\lambda}^{\rm{prim}}(k)Y(\eta,k)=\frac{v(k,\eta)}{a(\eta)}\, ,
\end{equation}
then the energy density of GWs is given by \cite{Caprini:2018mtu,Watanabe:2006qe}
\begin{equation}
\rho_{\rm{GW}}(\eta)=\frac{M_{\rm{Pl}}^2}{32\pi a(\eta)^2} \left<h_{ij}^{\prime}(k,{\bf x})h^{ij \prime}(k,{\bf x})\right>\, ,
\end{equation}
\begin{equation}
\label{haveg}
\left<h^\prime_{ij}(k,{\bf x})h^{ij\prime}(k,{\bf x})\right>=\int\frac{dk}{k}\mathcal{P}_T(k,\eta)\, .
\end{equation}
The tensor power spectrum is defined by
\begin{eqnarray}
\label{power0}
\mathcal{P}_T(k,\eta)&=&\frac{k^3}{\pi^2}\sum_{\lambda}\left<|h_{\lambda}(k,\eta)|^2\right> \nonumber \\
&=&\mathcal{P}^{\rm{prim}}_T(k)[Y(k,\eta)]^2\, ,
\end{eqnarray}
where its time independent part reads
\begin{equation}
\label{powinf}
\mathcal{P}^{\rm{prim}}_T(k)=\frac{k^3}{\pi^2}\sum_{\lambda}\left<|h^{\rm{prim}}_{\lambda}(k)|^2\right>
=\frac{1}{\pi}\left(\frac{4H_{\rm{inf}}}{M_{\rm{Pl}}}\right)^2\, .
\end{equation}
The Hubble parameter at the inflation scale is fixed by $H_{\rm{inf}}^2\approx (8\pi/3M_{Pl}^2)V_{\rm{inf}}$. 
The relic density of GWs can be obtained by
\begin{equation}
\label{relicgw}
\Omega_{\rm{GW}}(k,\eta)=\frac{\mathcal{P}^{\rm{prim}}_T(k)}{12 a(\eta)^2 H(\eta)^2} [Y'(k,\eta)]^2\, .
\end{equation}
Equations~(\ref{powinf}) and (\ref{relicgw}) show that the absolute value of the
relic density of PGWs depends on the inflationary scale.
Assuming an inflationary scale of $V_{\rm inf}^{1/4}=1.5\times10^{16}$~GeV the
relic density of GWs for the frequency range
between $10^{-9}$--$10^{-10}$~Hz will be $\Omega_{\rm{GW} }h^2\sim 2.4\times10^{-16}$
\cite{Saikawa:2018rcs}.

At the horizon it can be found that $[Y'(k,\eta)]^2=k^2[Y(k,\eta)]^2$. 
This can change when modes come well inside the horizon \cite{Watanabe:2006qe,Saikawa:2018rcs}.
Since our goal is to evaluate the effect of the EoS on the PGWs
and to compare their relic amplitude at high frequency to their relic amplitude at low 
frequency, using eqs.~(\ref{ampvh}), (\ref{powinf}), and (\ref{relicgw}), 
we can write
\begin{equation}
\label{gwrelicprop}
\Omega_{\rm{GW}}(k,\eta_0)\propto \Omega_{\rm{GW}}(k,\eta_{\rm{hc}})\propto k^5 |v(k,\eta_{\rm{hc}})|^2\, ,
\end{equation}
where the horizon crossing mode can be identified by 
\begin{equation}
\label{horcros}
k = a(\eta_{\rm{hc}})H(\eta_{\rm{hc}})\, .
\end{equation}
The temperature at horizon crossing can also be determined by using
eqs.~(\ref{fried2}) and (\ref{horcros}). One can also find the following
approximate relation between PGW relic, energy, and entropy
density at horizon crossing
\cite{Schwarz:1997gv,Watanabe:2006qe,Saikawa:2018rcs}
\begin{equation}
\label{gwrelicprop2}
\Omega_{\rm{GW}}(k,\eta_0)\propto \rho_{\rm{tot}}(T_{hc})s_{\rm{tot}}(T_{\rm{hc}})^{-4/3}\, .
\end{equation}
We can solve eqs.~(\ref{pgw-tensor}), (\ref{fried1}), and (\ref{fried2}) with
the initial conditions given by eq.~(\ref{initcon}) until the scale factor at horizon
reaches a value where each mode $k$ crosses the horizon or until a scale
factor at lower temperatures, e.g., after neutrino decoupling. After neutrino
decoupling, since the GWs evolve like radiation ($\rho_{\rm{GW}}\propto a^{-4}$) in
case of the absence of any phase transition afterwards, the spectrum will be
unchanged until today except for the damping in the amplitude due to the
expansion. This helps us to pin down the relative difference of PGWs for
different modes due to the evolution of the EoS with temperature in
the early universe. We do not consider the effect of an anisotropic stress
due to the free streaming of photons and decoupled neutrinos, which appears as
a source on the right hand side of eq.~(\ref{tensor-pert}), because it is
effective only for frequencies smaller than $\sim 5\times10^{-11}$~Hz
\cite{Weinberg:2003ur,Watanabe:2006qe,Saikawa:2018rcs}.


\section{\label{sec:qcdeos}The Role of the Equation of State of the SM on PGWs}

In order to solve eq.~(\ref{pgw-tensor}) for different GW wave numbers $k$ it is required to first solve eqs.~(\ref{fried1}) and (\ref{fried2}) to find the temperature as
a function of $a$ or $\eta$. For this purpose, we should know the quantities
$\rho_{\rm{tot}}$ and $p_{\rm{tot}}$ at each temperature $T$. The total energy and
pressure density can be computed from the following equations
\footnote{The energy and entropy density can also be written as a function of
  degrees of freedom,
  i.e.\ $\rho_{\rm{tot}}(T,\mu)=(\pi^2/30)g_{\rho,\rm{tot}}(T,\mu(T))T^4$ and
  $s_{\rm{tot}}(T,\mu)=(2\pi^2/45)g_{s,\rm{tot}}(T,\mu(T))T^3$.}
\begin{eqnarray}
\label{energy-density}
\rho_{\rm{tot}}(T,\mu)&=& \sum_{i} \frac{g_i}{2\pi^2}\int_{m_{i}}^{\infty}dE\times
                       \nonumber \\
                 && E^2\sqrt{E^2-m_{i}^2} \left(\frac{1}{e^{\frac{E-\mu_{i}}{T}}\pm1}\right)\, ,\\
\label{pressure-density}
p_{\rm{tot}}(T,\mu)&=& \sum_{i} \frac{g_i}{6\pi^2}\int_{m_{i}}^{\infty}dE\times
                                        \nonumber \\
                 && \left(E^2-m_{i}^2\right)^{3/2} \left(\frac{1}{e^{\frac{E-\mu_{i}}{T}}\pm1}\right)\, ,
\end{eqnarray}
with the sum over all particle species $i$ with degrees of freedom $g_i$ and chemical potential $\mu_i$.
The total entropy density is given by
\begin{eqnarray}
\label{eq:entropytot}
  Ts_{\rm{tot}}(T,\mu)&=&\rho_{\rm{tot}}(T,\mu)+p_{\rm{tot}}(T,\mu) \nonumber \\
                 &&-\sum_{i}\mu_{i}n_{i}(T,\mu_{i}),
\end{eqnarray}
For each particle species the net number density of particles minus
anti--particles can be defined as
\begin{eqnarray}
\label{number-density}
n_{i}(T,\mu_{i})&=&\frac{g_i}{2\pi^2}\int_{m_{i}}^{\infty}dE~E\sqrt{E^2-m_{i}^2}
                    \times \nonumber \\
&&\left(\frac{1}{e^{\frac{E-\mu_{i}}{T}}\pm1}-\frac{1}{e^{\frac{E+\mu_{i}}{T}}\pm1}\right)\, .
\end{eqnarray}
The above equations can be used to determine the energy and entropy density of
the SM which can be implemented in eqs.~(\ref{pgw-tensor}), (\ref{fried1}), and
(\ref{fried2}) to compute the relic density of PGW in the
early universe according to eq.~(\ref{relicgw}).

In this section we only consider the case of vanishing chemical potentials.
Several studies have been performed before for the case of vanishing chemical
potentials using different lattice QCD results around the QCD 
transition available at that time
\cite{Borsanyi:2016ksw,Laine:2006cp,Drees:2015exa,Laine:2015kra,Saikawa:2018rcs,Hindmarsh:2005ix}
and considering different assumptions for the EoS in the
perturbative regime of QCD, at the electroweak transition, and for neutrino
decoupling.
 
As shown in refs.~\cite{Schettler:2010dp,Saikawa:2018rcs} the characteristic
frequency of PGWs related to the QCD transition temperature, $T_{\rm{QCD}}\sim 150$~MeV,
is $f_{\rm{QCD}}\approx 3\times10^{-9}$~Hz. The effect of neutrino decoupling at
low temperature ($T\sim1$~MeV and $f\sim10^{-11}$~Hz) is important to compute
the precise value of temperature with respect to the scale factor
\cite{Lesgourgues:2012uu}. This effect appears due to the varying temperatures
of different neutrino flavors during and after neutrino decoupling
\cite{Lesgourgues:2012uu}.

\begin{figure}
\includegraphics[scale=0.56]{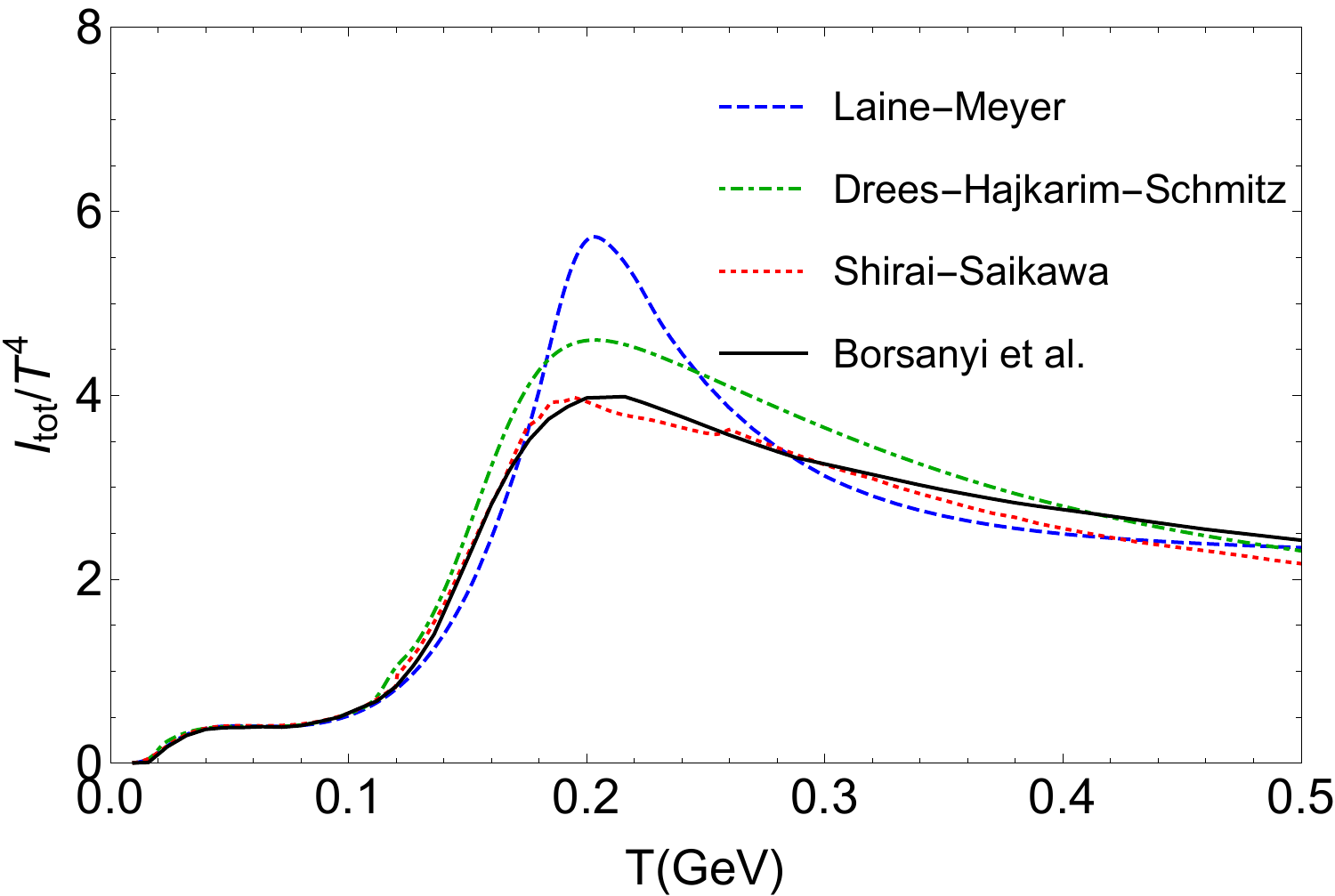}
\caption{\label{trace} The trace anomaly including lattice QCD results
  for temperatures up to $500$~MeV taken from different approaches
  \cite{Laine:2015kra,Saikawa:2018rcs,Drees:2015exa,Borsanyi:2016ksw}. See
  text for details.}
\end{figure}

\begin{figure}
\includegraphics[scale=0.55]{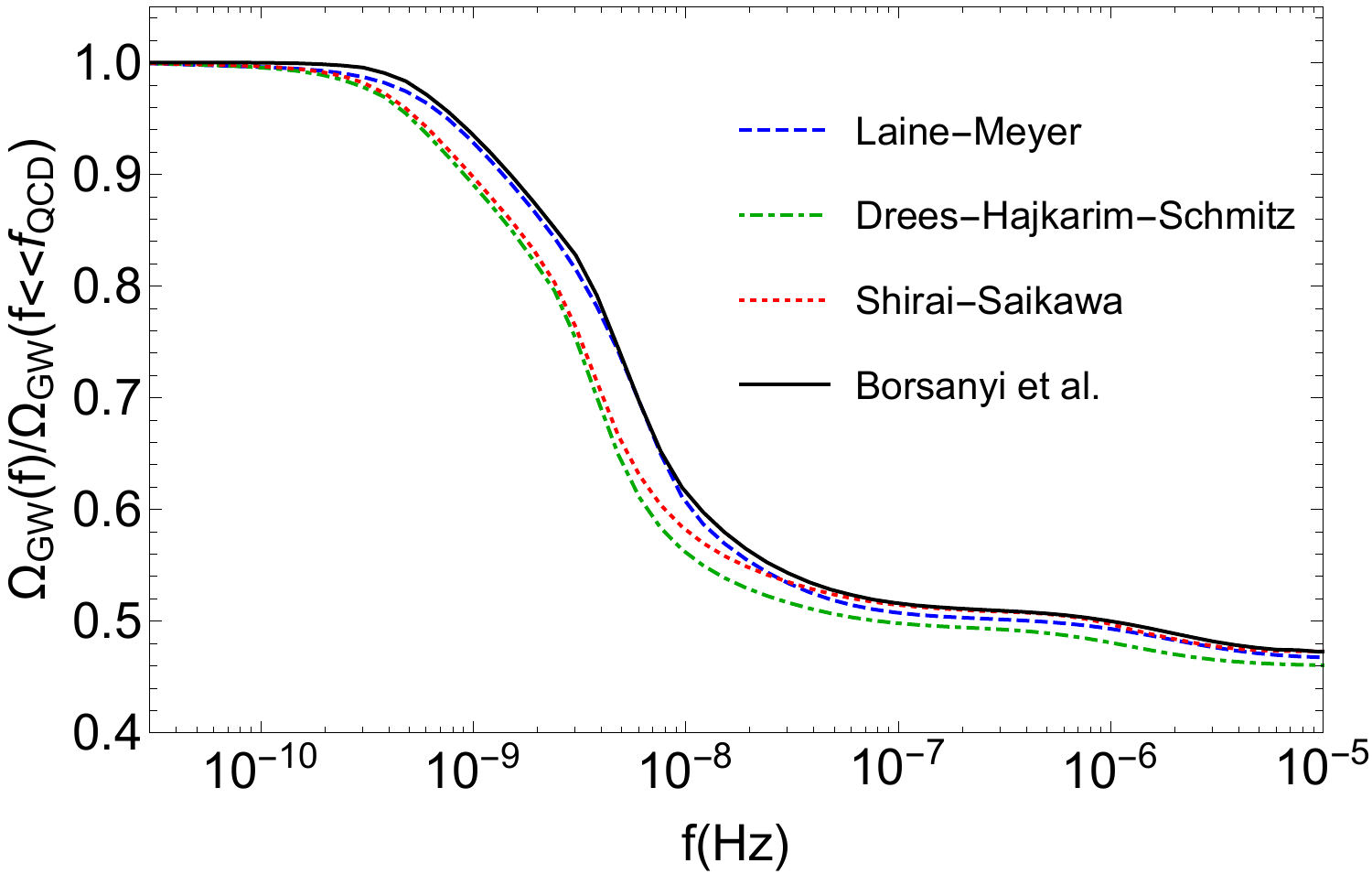}
\caption{\label{gw-spectrum} The PGW spectrum with
  respect to frequency using different EoSs for the early universe shown in
  the previous figure.}
\end{figure}
 
The main QCD EoS we use is the one by ref.~\cite{Borsanyi:2016ksw} (labeled as
`Borsanyi et al.' in the figures). In ref.~\cite{Borsanyi:2016ksw} the
EoS from $10$~MeV to ($500$~MeV) $1$~GeV for ($2+1$) $2+1+1$ flavors by using
lattice methods is computed. Their computed trace anomaly
for all SM particles and the predicted PGW spectrum are shown in
fig.~\ref{trace} and \ref{gw-spectrum}, respectively. By using hard thermal
loop corrections and the perturbative QCD approach up to order
$\mathcal{O}(g_s^6)$ including effects from charm and bottom quarks,
they derive the EoS for temperatures above $1$~GeV
\cite{Andersen:2010wu,Kajantie:2002wa}. They use the results of
ref.~\cite{Laine:2015kra} for very high temperatures around the electroweak transition. 
For smaller temperatures they adopt a hadron resonance gas
(HRG) model approach \cite{Huovinen:2009yb}. Thereby they provide the EoS for 
temperatures between $1$~MeV and $100$~GeV including all SM
particles. The data set of ref.~\cite{Borsanyi:2016ksw} is the one we use to
extract the impact of charm quarks from lattice QCD calculations on PGWs.

In fig.~\ref{trace} and fig.~\ref{gw-spectrum} the corresponding result of
\cite{Drees:2015exa} for the QCD EoS of the standard model is shown using the
data of the HotQCD Collaboration \cite{Bazavov:2014pvz} for $2+1$ flavors and the
EoS including also charm quark by ref.~\cite{Borsanyi:2010cj}. Also, the HRG model data
is used for temperatures between $70$ and $100$~MeV
\cite{Huovinen:2009yb}. The remaining SM particles are assumed to be free
particles. Moreover, the effect of neutrino decoupling has been considered
using ref.~\cite{Lesgourgues:2012uu}. The interpolated result for the EoS of
ref.~\cite{Bazavov:2017dus}, using another lattice calculation for vanishing
chemical potentials, highly matches the result of \cite{Bazavov:2014pvz} in the
temperature range of $130-230$~MeV, so that we do not include their results in
this paper.

Additionally, we compare the thermal evolution 
of the EoS of the SM with a calculation by
Laine and Meyer \cite{Laine:2015kra} who used the older treatment of Laine and
Schroeder \cite{Laine:2006cp} for temperatures below $110$~GeV. In
ref.~\cite{Laine:2006cp} the radiative corrections up to order 
$\mathcal{O}(g_s^2)$ for a running quark mass are considered. For temperatures
below $350$~MeV old lattice data for pure glue theory is used. The EoS of ref.~\cite{Laine:2006cp}
includes temperatures between $10$~MeV and $1$~TeV including all SM
particles. However, since the mass of the Higgs boson was unknown at that
time, they have considered a different value from the nowadays accepted
one. This issue is fixed in the work of Laine and Mayer \cite{Laine:2015kra}
by considering $125$~GeV for the Higgs mass and including corrections from
lattice field theory calculations around the electroweak transition
($110$~GeV$\lesssim T\lesssim 250$~GeV). For temperatures above $250$~GeV they
have considered the perturbative result of \cite{Gynther:2005dj}. The result for their
treatment for the trace anomaly up to a temperature of $500$~MeV is
shown in fig.~\ref{trace}. The comparison of the influence of the EoS 
of Laine and Meyer on the PGW spectrum with the other EoSs for 
the early universe is shown in fig.~\ref{gw-spectrum}.

Another treatment of the SM EoS that we investigate is from the work of
ref.~\cite{Saikawa:2018rcs} which used lattice QCD results for $2+1+1$ 
flavors \cite{Borsanyi:2016ksw} matched to a HRG model
\cite{Huovinen:2009yb} at temperatures below the QCD transition and
perturbative QCD up to $\mathcal{O}(g_s^6\log g_s)$ at higher temperatures
\cite{Kajantie:2002wa}. For temperatures around the electroweak transition the
results of refs.~\cite{Laine:2015kra,DOnofrio:2015gop,Buttazzo:2013uya}
considering lattice calculations and perturbative calculations are used. At
very low temperatures around electron and neutrino decoupling the result of
\cite{deSalas:2016ztq} is adopted to consider the effect of neutrino
oscillations. The effect of this EoS on the trace anomaly, PGW and the differences compared to
result obtained with the EoS of ref.~\cite{Borsanyi:2016ksw} are shown in figs.~\ref{trace} and
\ref{gw-spectrum}, respectively.

In fig.~\ref{trace} results for the trace anomaly of QCD based on different
lattice calculations reported in the literature are shown. It can be seen that 
the location of the peak in the trace anomaly is similar but the
height differs for different approaches as different input from lattice
calculations have been used. Figure \ref{gw-spectrum} shows the effect of
using the different results for the trace anomaly 
on the relic density of PGWs as a function of the frequency. For frequencies in the range
$10^{-10}-10^{-9}$~Hz ($T\sim 10-100$~MeV for horizon crossing) the HRG
model and muons play the major role for the PGW relic density. For frequencies
between $10^{-9}$ and $10^{-7}$~Hz the EoS including lattice QCD results, tau
leptons and bottom quarks are important for determining the GW relic density
for horizon crossing temperatures between $100$~MeV and $10$~GeV. For
temperatures around $100$~GeV and frequencies around $10^{-6}$~Hz the
appearance of top quarks and the electroweak sector including W$^\pm$, Z, and Higgs
bosons have the dominant impact on the SM EoS and the prediction of the stochastic
GW background.

\begin{figure}
\includegraphics[scale=0.56]{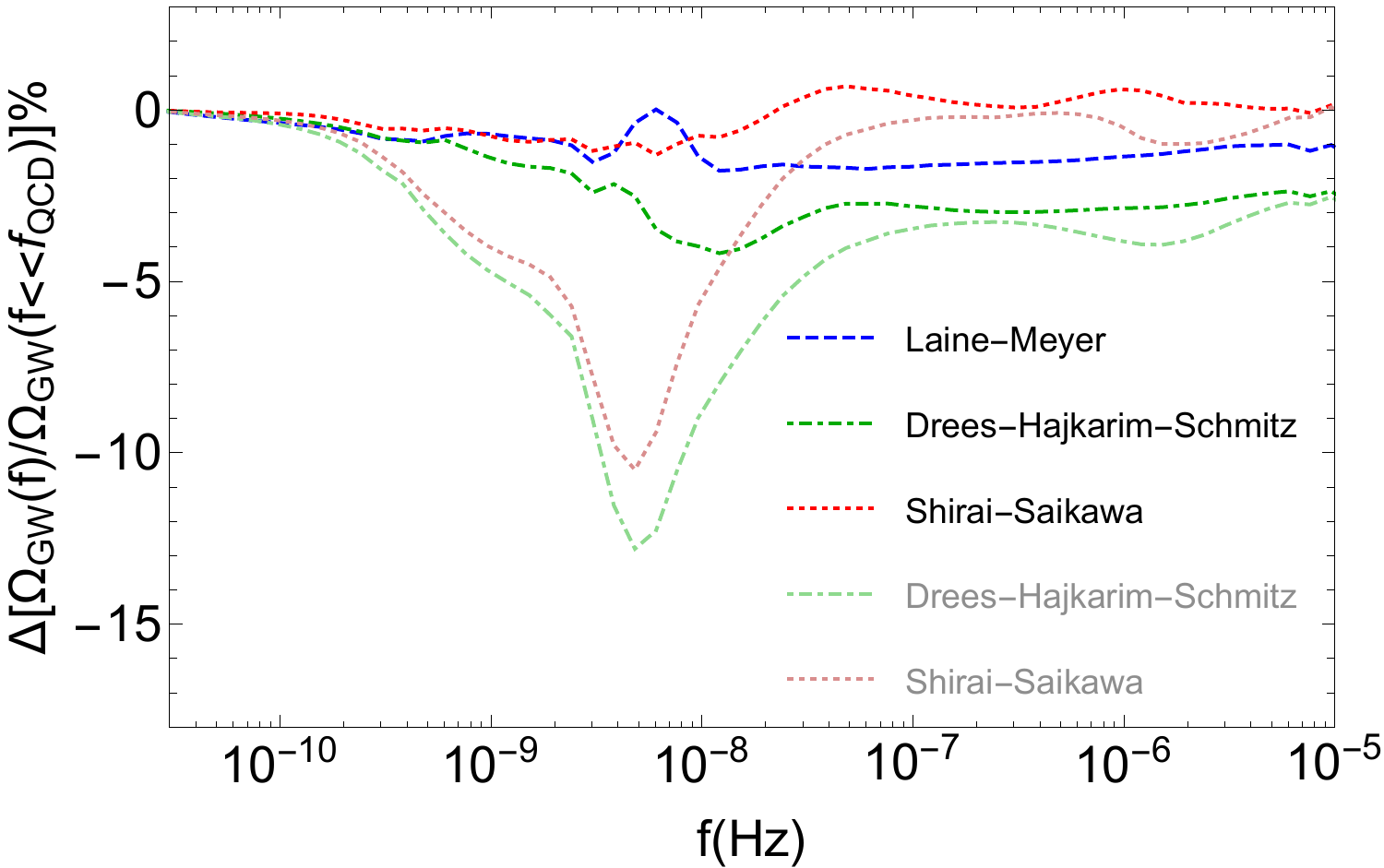}
\caption{\label{wupperother} The relative differences of the predicted
  relic density of PGWs using the EoSs from
  refs.~\cite{Laine:2015kra,Saikawa:2018rcs,Drees:2015exa} compared to the one of
  ref.~\cite{Borsanyi:2016ksw}. The opaque curves do
  not include the effect of electron and neutrino decoupling to better compare the role of
  QCD on the relic density of PGWs with the main data set \cite{Borsanyi:2016ksw}. 
  The transparent curves include the contribution due to electron and neutrino decoupling.}
\end{figure}
 
Two of the EoSs \cite{Drees:2015exa,Saikawa:2018rcs} shown in
fig. \ref{gw-spectrum} consider effects from neutrino decoupling but this effect is not
taken into account for the
other EoSs \cite{Borsanyi:2016ksw,Laine:2015kra}. 
Neutrino decoupling leads to a shift of the PGW
spectrum to higher frequencies compared to the case without considering neutrino
decoupling taken into account. Moreover, it causes a relative error of up to
$\sim10$\% between two cases as shown in fig.~\ref{wupperother}. The
discrepancy originates mainly from the change in determining the precise
relation between the scale factor and the temperature using
eqs.~(\ref{fried1}) and (\ref{fried2}) which differs if one takes into account
neutrino decoupling or not. The differences in the trace anomaly (shown in
fig.~\ref{trace}), energy, and entropy density are mostly due to the
various treatments of the QCD EoS. As fig.~\ref{wupperother} shows 
using different EoSs results in a deviation of up to $3-4$\% at
frequencies around $3\times10^{-9}$~Hz and up to $3$\% for higher frequencies
if one neglects the effect from neutrino decoupling.  
The deviations in the predicted PGW relic shown in fig.~\ref{gw-spectrum} 
are computed at the scale factor of horizon crossing for each mode which is 
numerically doable. We also studied the deviations for a limited frequency
 range ($3\times10^{-11}$ and $5\times10^{-8}$ Hz) at a fixed scale factor after 
neutrino decoupling when the evolution of the EoS will not be affected by 
SM particles any more. Our results show that highly evolving 
the EoS especially around the QCD transition improve the predicted  PGW relic 
around $1$\%, since the deviation of the EoS from radiation
causes a small damping of PGWs after horizon crossing. We do not
show the plot for this calculation in this paper, since our frequency range
 is limited and doing it for a larger frequency range is numerically expensive.
These discrepancies
between different treatments of the EoS can be distinguished by SKA, EPTA, LISA,
and DECIGO at different frequencies by the observation of PGWs.

\begin{figure}
\includegraphics[scale=0.54]{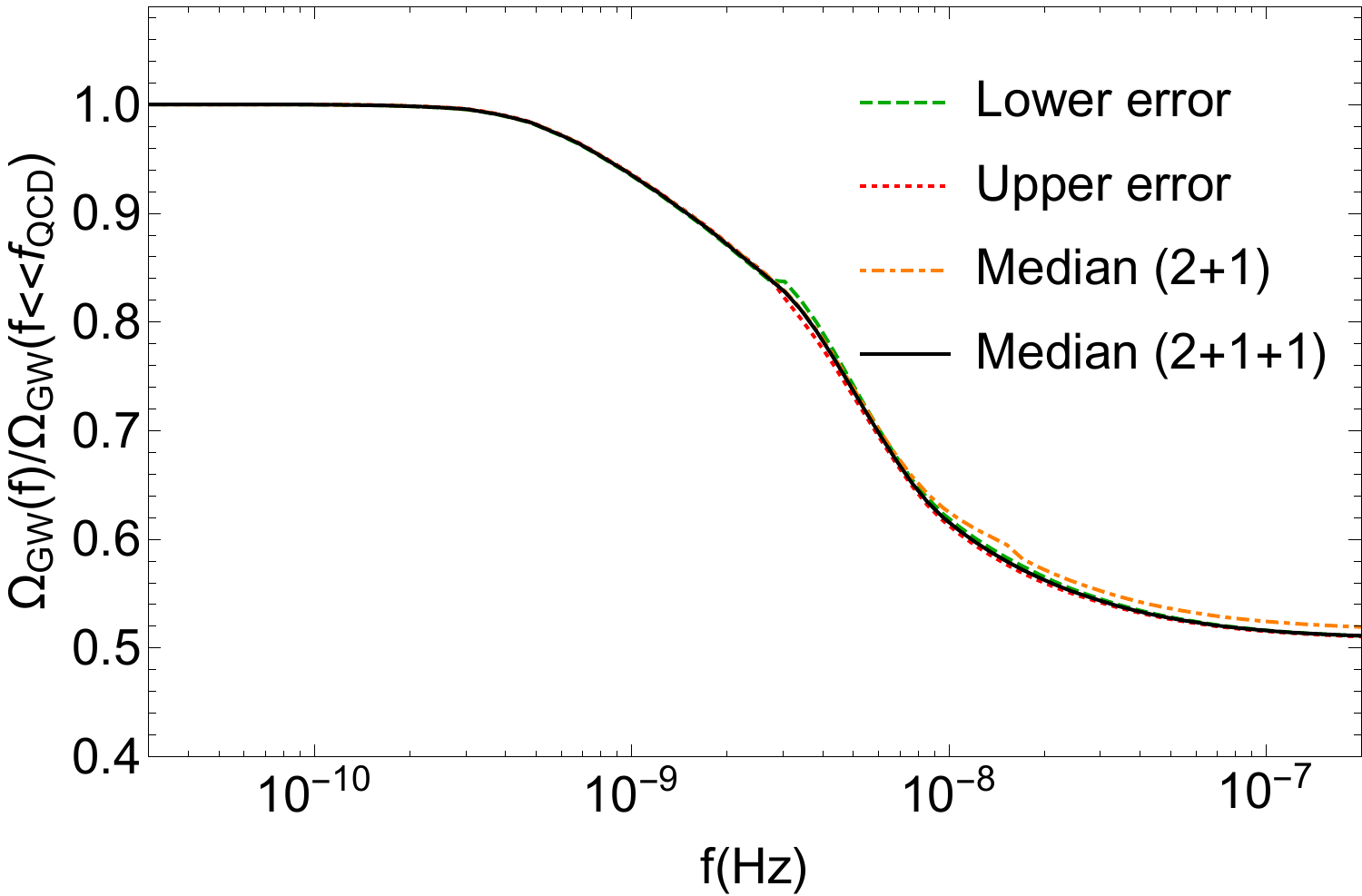}
\caption{\label{wupperrorband} The error band of the relic density of PGWs from
  median value ($2+1+1$) by taking the lower and upper bounds from lattice
  data of ref.~\cite{Borsanyi:2016ksw}. Also, the effect of including charm
  quarks in lattice simulations on the he relic density is presented using the data of
  ref.~\cite{Borsanyi:2016ksw} for $2+1$ and $2+1+1$ flavors for the median
  values. As it can be seen there is a deviation between the computed relic density 
  of PGWs due to charm quarks mostly for frequencies around $10^{-8}$~Hz and
  higher.}
\end{figure}

\begin{figure}
\includegraphics[scale=0.39]{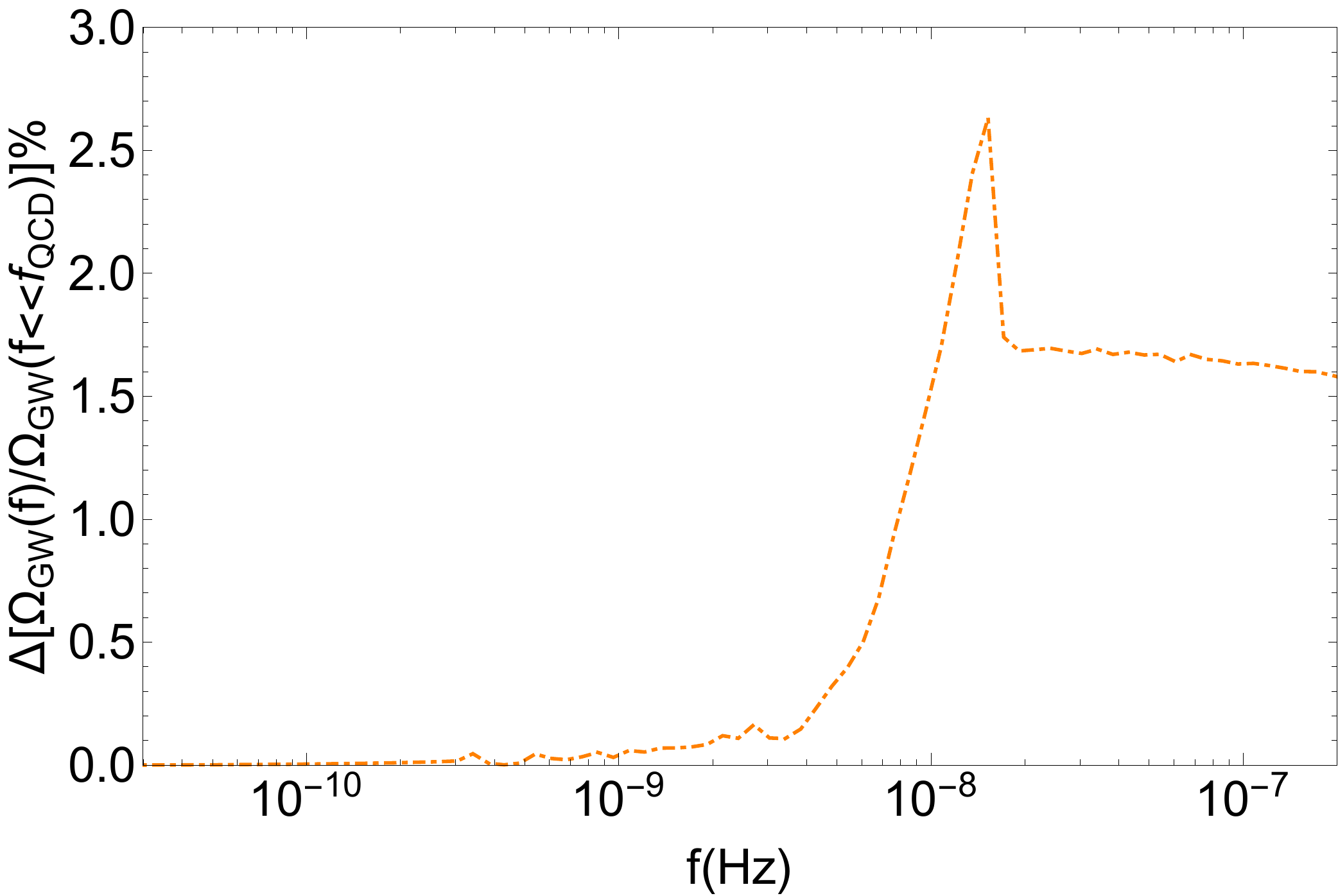}
\caption{\label{charmed} The error due to the effect of charm quarks at low
  temperatures in lattice simulations on PGWs is presented using
  the data of ref.~\cite{Borsanyi:2016ksw} for $2+1$ and $2+1+1$ flavors.}
\end{figure}

In fig.~\ref{wupperrorband} the effect of considering charm quarks in lattice
calculations on the predicted relic density of PGWs is shown using lattice data
from ref.~\cite{Borsanyi:2016ksw} for $2+1$ and $2+1+1$ flavors. The
difference is small but not negligible. The relative difference of the
predicted pattern of PGWs between $2+1$ and $2+1+1$ flavors in lattice QCD for
different frequencies is shown in fig.~\ref{charmed} and amounts up to $2.6$\%.
For frequencies higher than $2\times10^{-8}$~Hz the difference is due to the
change in the relation between energy density and 
scale factor computed from eq.~\ref{fried2}, since a lower
temperature as an initial
condition affects this relation for the higher temperature.

\begin{figure}
\includegraphics[scale=0.56]{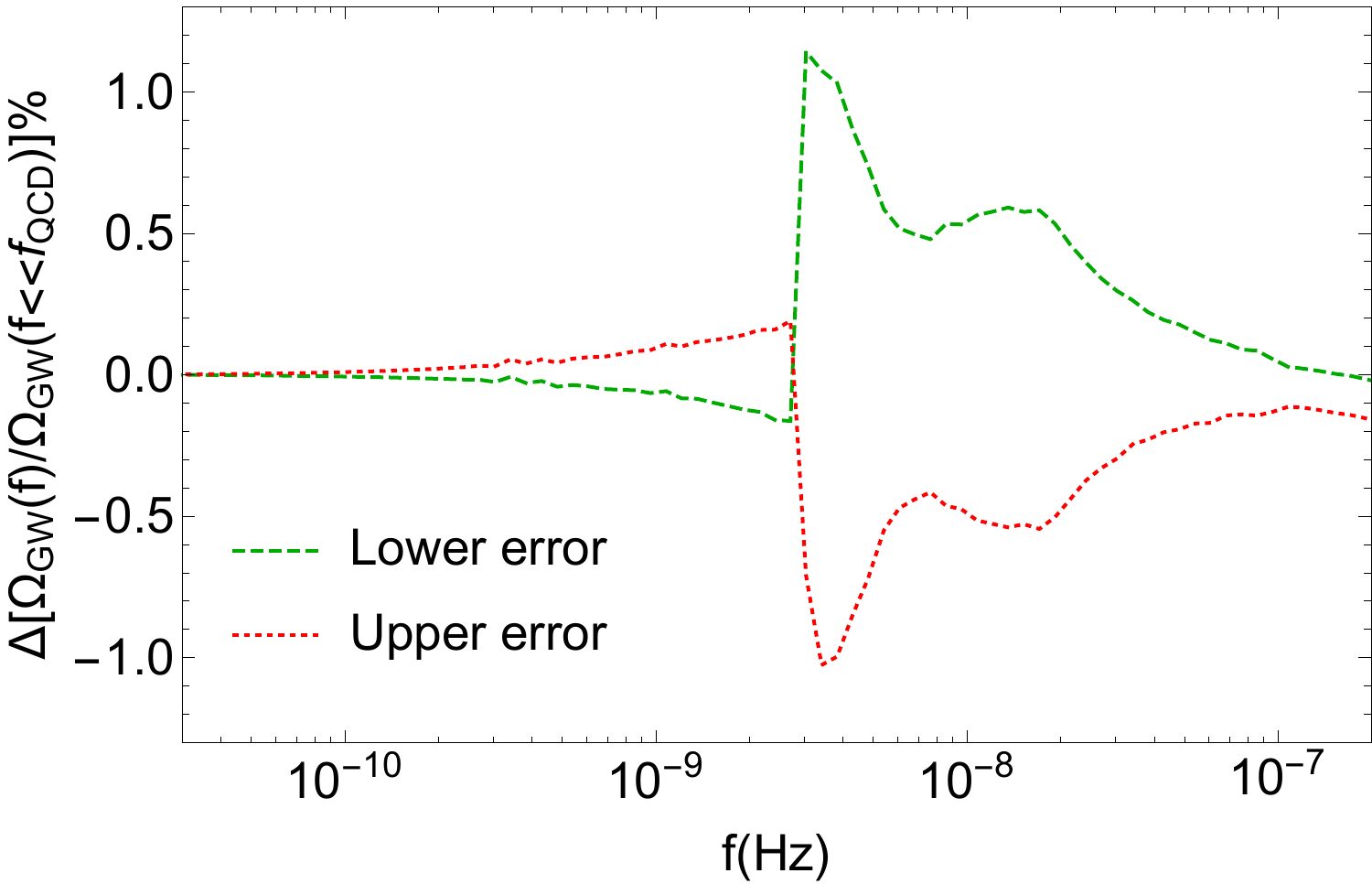}
\caption{\label{wupperror} The relative error on the relic density of PGWs due to
  the uncertainties in the EoS from lattice QCD and perturbative calculations
  from ref.~\cite{Borsanyi:2016ksw}.}
\end{figure}

We also consider the uncertainties
of the lattice data and the perturbative QCD calculations up to
$\mathcal{O}(g_s^6)$ for the temperature range of $100$~MeV to $10$~GeV
considering $2+1+1+1$ flavors, i.e., including effects from bottom
quarks, which is discussed in ref.~\cite{Borsanyi:2016ksw}. 
The resulting error band in the relic density of PGWs is depicted in
fig.~\ref{wupperrorband}. One sees that the changes are small. In fact, the
relative error in the relic density of PGWs amounts to a discrepancy of at most
$1.1$\% around the QCD transition as shown in fig.~\ref{wupperror}.


\section{\label{sec:chempot}SM with Nonvanishing Chemical Potentials and PGWs}

The value of the lepton asymmetry in the universe is constrained by analyses of BBN and the 
cosmic microwave background (CMB) to be \cite{Oldengott:2017tzj}:
\begin{equation}
\label{lepasym}
l=\frac{n_L}{s}\lesssim0.012\, .
\end{equation}
Also from the Planck data \cite{Aghanim:2018eyx} one knows the amount of baryon
asymmetry of the universe
\begin{equation}
\label{basym}
b=\frac{n_B}{s}\approx 8 \times10^{-11}\, .
\end{equation}
Considering SM particles in a thermal bath and assuming that sphaleron
processes occur efficiently then the lepton asymmetry is related to the baryon
asymmetry by $l=-\frac{51}{28}b$ \cite{Harvey:1990qw}. Such tiny values 
for the lepton asymmetry and baryon asymmetry do not lead to a first order phase
transition of QCD \cite{Wygas:2018otj,Bazavov:2018mes}. 
However, such a small value
of $l$ has not been confirmed experimentally. 
The effect of a sizable lepton asymmetry on the evolution of the chemical potentials of SM
particles with respect to temperature and its effect on the cosmic trajectory
has been investigated in refs.~\cite{Schwarz:2009ii,Wygas:2018otj}.

In the early universe, between neutrino oscillations ($T_{\rm{osc}}\sim 10$~MeV) 
and the electroweak transition ($T_{\rm{ew}}\sim 100$~GeV), conservation 
of nonvanishing lepton flavor asymmetries, baryon
asymmetry, and electric charge leads to the following set of equations
\cite{Schwarz:2009ii,Wygas:2018otj}:
\begin{eqnarray}
l_f s(T,\mu)&=& n_f(T,\mu_{f})+n_{\nu_f}(T,\mu_{\nu_f})\, , f=e, \mu, \tau \, , \nonumber \\
bs(T,\mu) &=& \sum_{i}b_i n_i(T,\mu_{i})\, , \nonumber \\
0 &=& \sum_{i}{q_i n_i(T,\mu_{i})}\, ,
\end{eqnarray}
with $n_{i}$ is the net number density of particles minus anti-particles 
given by eq.~(\ref{number-density}) and we presumed electric charge neutrality of the universe in the last equation.

The total entropy density can be determined according to eq.~\eqref{eq:entropytot} using eqs.~\eqref{energy-density} and \eqref{pressure-density} considering all relevant SM particles and their chemical potentials. Solving this system of coupled equations at a given temperature we get the temperature evolution of the SM chemical potentials \cite{Wygas:2018otj} and thus we can compute the total pressure and energy density for nonvanishing lepton asymmetries (cf.~\cite{Wygas:2018phd}). For the numerical evaluation we assumed equally distributed lepton flavor asymmetries, $l_f=l/3$.

For different temperature ranges one can find approximate relations between
the lepton flavor chemical potentials and the electric charge chemical potential. 
For example in
the temperature range where lattice QCD plays a major role, i.e.,
$150$~MeV$<T<350$~MeV (the upper bound is defined due to the presence of charm quarks
at higher temperatures) we have $\mu_Q\approx\mu_{l_f}/2$. For temperatures between
the QCD transition and the temperature where pions are not 
relativistic anymore ($m_{\pi^{\pm}}/3<T<150$~MeV) one finds
$\mu_Q\approx2\mu_{l_f}/3$. For even lower temperatures but above neutrino decoupling one
can find $\mu_Q\approx \mu_{l_f}$ due to charge conservation
\cite{Schwarz:2009ii,Wygas:2018otj,Gebhardt:2018}.

\begin{figure}
\includegraphics[scale=0.54]{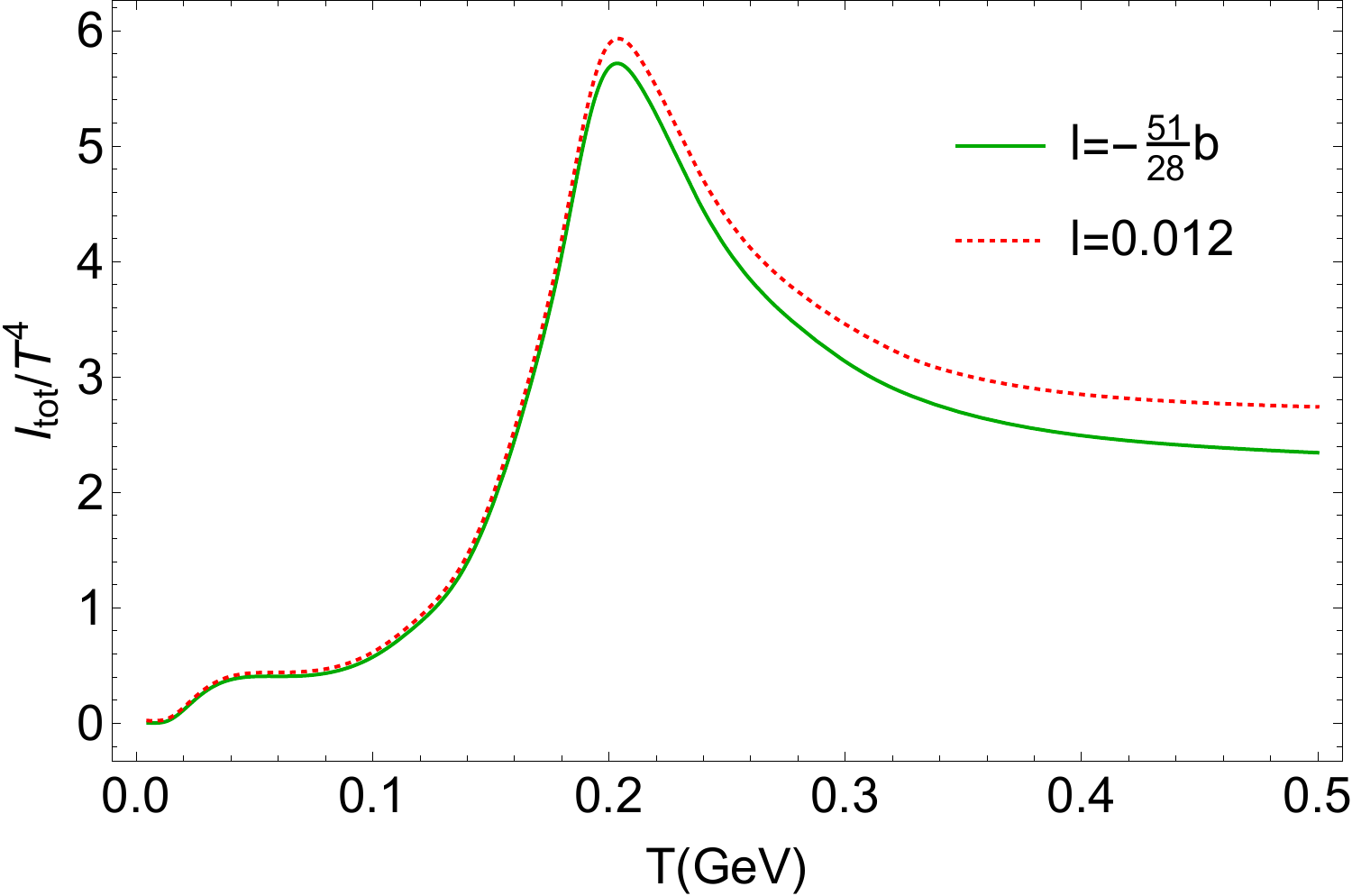}
 \caption{\label{trace-lep} The trace anomaly for vanishing and nonvanishing lepton
  asymmetry.
  }
\end{figure}

\begin{figure}
\includegraphics[scale=0.55]{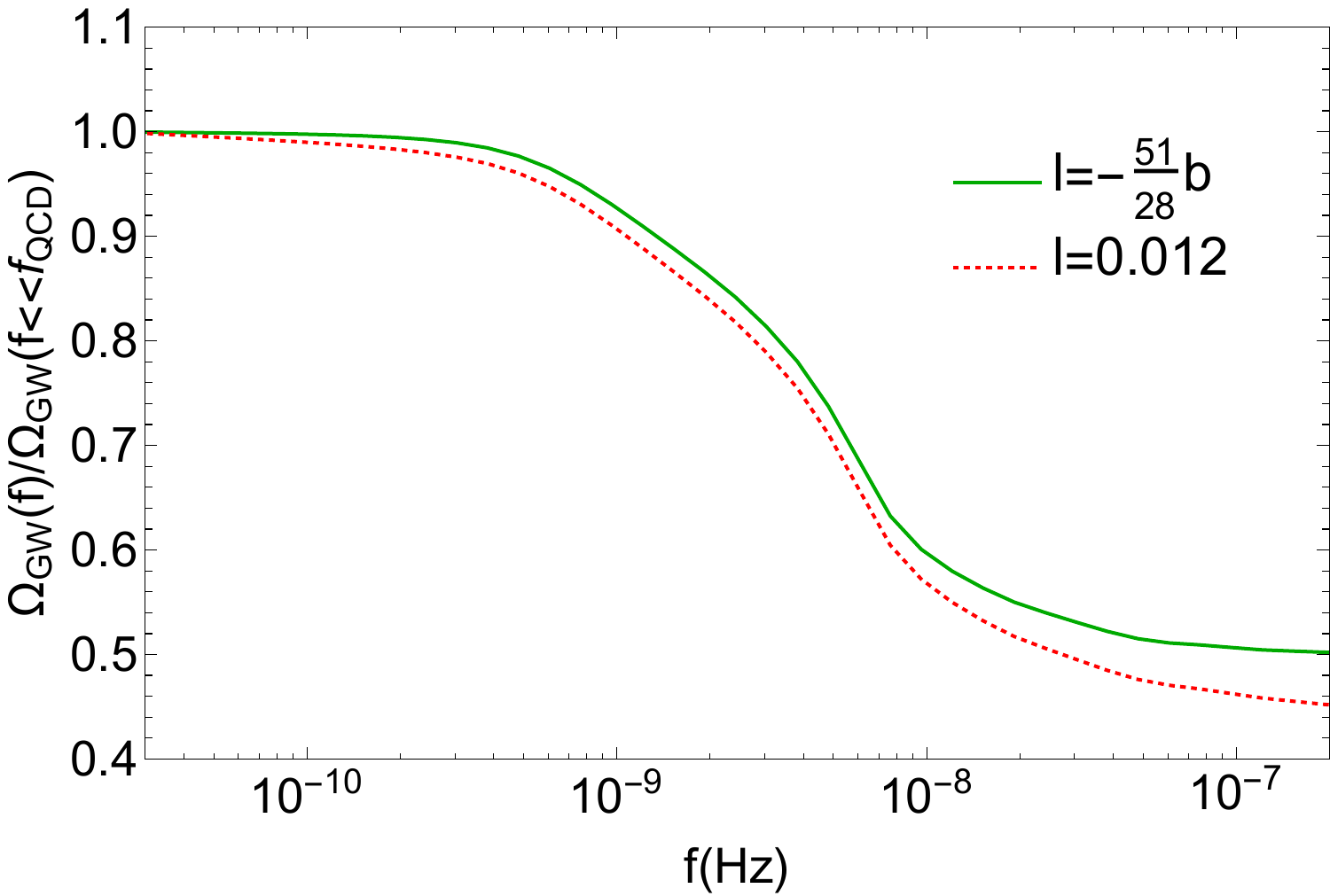}
\caption{\label{gw-spectrum-lep} The scaled relic density of PGWs versus
  frequency assuming different values for the lepton asymmetry.
  }
\end{figure}

To calculate the influence of a nonvanishing lepton asymmetry on PGWs one needs
the temperature evolution of the energy and entropy density at nonvanishing lepton asymmetry for the early
universe. We calculate the EoS between 150 MeV and 350 MeV 
according to ref.~\cite{Wygas:2018otj, Wygas:2018phd} using lattice QCD 
susceptibilities ($\chi_{ab}=(\partial^2 p/\partial \mu_a \partial \mu_b)\vert_{\mu=0}$) 
\cite{Bazavov:2014yba,Mukherjee:2015mxc} to determine the evolution of the 
chemical potentials at nonvanishing lepton asymmetry.
The hadron resonance gas model (HRG) is computed by using a similar approach as in
\cite{Huovinen:2009yb} for temperatures below $100$~MeV by considering hadrons
up to a mass of $2.5$~GeV as an ideal gas. 
The energy and entropy density at nonvanishing chemical potentials for
temperatures above 350~MeV are calculated as described before
according to \cite{Wygas:2018otj}, using the results of \cite{Laine:2006cp} 
for considering perturbative QCD effects up to order $\mathcal O (g_s^2)$ 
in case of vanishing chemical potentials.

\begin{figure}
\includegraphics[scale=0.4]{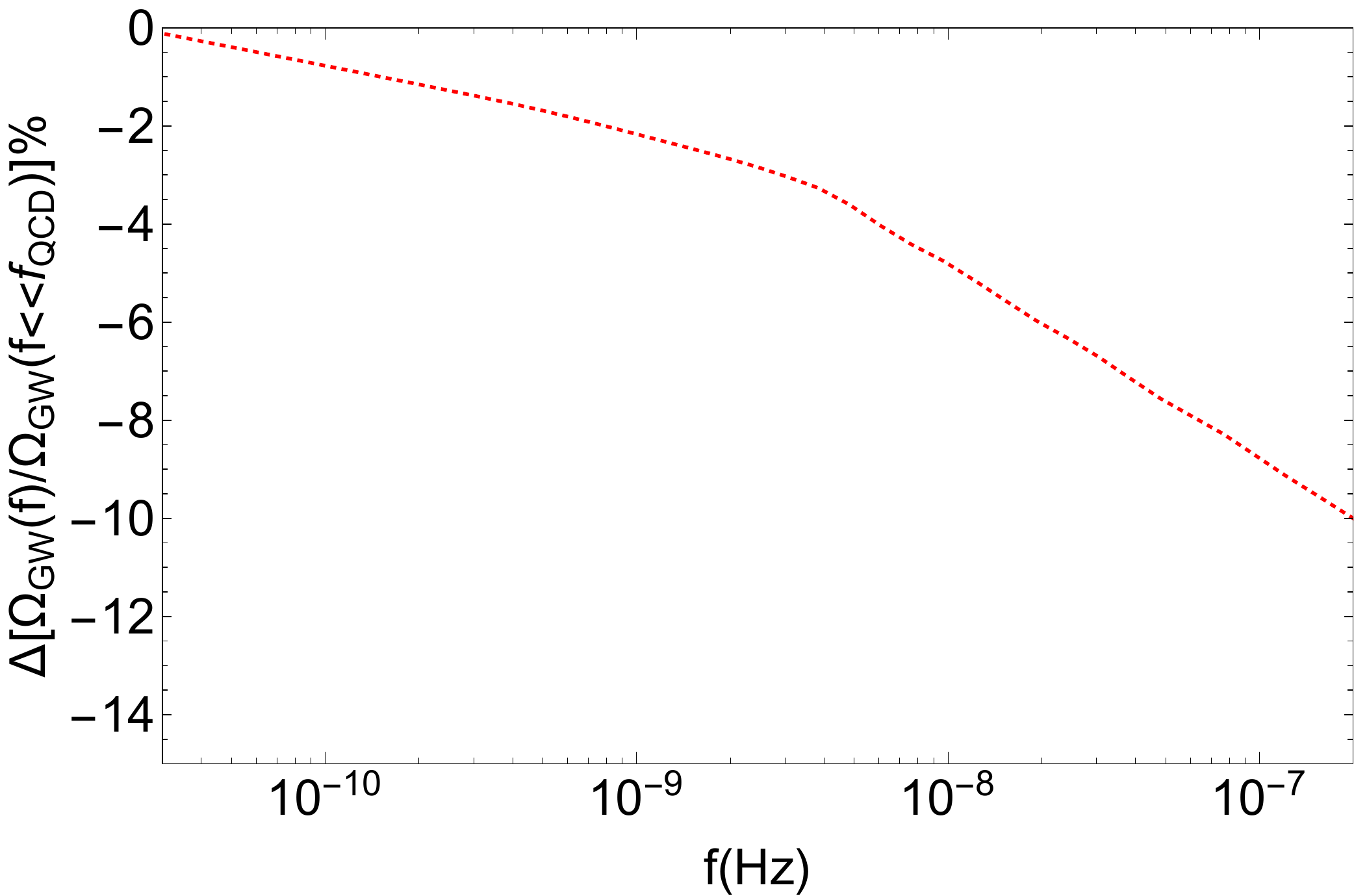}
\caption{\label{mandylep-gwdiff} The relative change of the relic density of PGWs 
  due to the presence of nonvanishing lepton chemical potentials for the SM in the thermal bath of
  the early universe using the data of fig.~\ref{gw-spectrum-lep}.}
\end{figure}

Figures~\ref{trace-lep} and \ref{gw-spectrum-lep} show the trace anomaly and the
relic density of PGWs for different values of the lepton asymmetry in the
early universe, respectively. There is up to a $10$\% difference between considering nearly
vanishing ($l=-\frac{51}{28}b$) and nonvanishing lepton asymmetry ($l=0.012$,
$l_{e}=l_{\mu}=l_{\tau}=0.004$). This result is based on the computation at horizon crossing. We also
 checked it with the calculation at a specific scale factor after neutrino 
 decoupling and found the difference between these two methods to be 
 less than $1$\%.
Here we used the EoS calculated according to \cite{Wygas:2018phd} 
in the PGWs relic density for frequencies above $\sim 10^{-11}$~Hz.
A deviation of the EoS from the predicted value for vanishing lepton asymmetry can be
measured in the spectrum of PGWs for frequencies around $10^{-9}-10^{-6}$~Hz by SKA or EPTA, at higher
frequencies $10^{-5}-10^{-2}$~Hz by LISA (this frequency range is outside
the range plotted in fig.~\ref{mandylep-gwdiff}), or at $10^{-3}-1$~Hz by
DECIGO. The detection of a sizeable lepton asymmetry in the early
universe can give impetus for possible scenarios for the explanation of
 the matter-antimatter asymmetry in the early universe and today. 
 We would like to emphasize that such a small
deviation in the EoS of the SM from a vanishing lepton asymmetry can not be observed by CMB
measurements, since its presence mostly shows up before BBN when more SM
particles are present in the thermal bath.

 
\section{\label{sec:conclusion}Summary and Conclusions}
 
In this paper we studied the effect of the QCD EoS using different
lattice QCD simulations including vanishing and nonvanishing chemical potentials
\cite{Bazavov:2014pvz,Laine:2015kra,Borsanyi:2010cj,Borsanyi:2016ksw,Wygas:2018otj}
on the relic density of PGWs produced by the inflationary
scenario. These kind of GWs can be observed by different experiments at a level
of less than $1$\% in the relic density per frequency depending on the length of
observation and the sensitivity
\cite{Seto:2001qf,Sato:2017dkf,Audley:2017drz,Janssen:2014dka,Lentati:2015qwp}.
The SKA and EPTA experiments are designed for frequencies
$10^{-9}-10^{-6}$~Hz by measuring the variation in the distance of pulsars. SKA 
can observe the PGW background relic density for values as small as
$\Omega_{\rm{GW}}h^2\approx10^{-16}$ depending on the time of exposure which is at
the order of few decades. Other experiments like LISA and DECIGO which are
proposed for larger frequencies $10^{-5}-1$~Hz can also probe the QCD effects
mostly due to perturbative effects on the EoS of the SM at higher temperatures.
 
Different sets for the EoS from
refs.~\cite{Laine:2015kra,Saikawa:2018rcs,Drees:2015exa,Borsanyi:2016ksw} have
been used to calculate the PGW spectrum. The difference between the various EoSs
of the SM, using different lattice QCD data as input, results in a relative
difference of up to $4$\% (fig.~\ref{wupperother}) in the relic
density of PGWs, mostly between frequencies of $10^{-9}-10^{-8}$~Hz and somewhat
smaller differences at higher frequencies. We also considered the uncertainties of the
EoS given in \cite{Borsanyi:2016ksw} by taking lower and upper error
bounds of lattice QCD calculations into account which leads to a deviation of up to $1.1$\%
(fig.~\ref{wupperror}) in the relic density of PGWs in the frequency range of
$10^{-9}-10^{-7}$~Hz. Additionally, we investigated the effect of considering
charm quarks using lattice QCD data at temperatures lower than $1$~GeV which
causes up to $2.6$\% deviation in the predicted PGW density
around a frequency of $10^{-8}$~Hz (fig.~\ref{charmed}).

We also discussed the effect of a nonvanishing lepton asymmetry on the
EoS of matter in the early universe. Our calculation shows that
this can lead to a difference of up to $2-10$\% in
the relic density of PGWs for frequencies around and larger than
$f_{\rm{QCD}}\sim3\times10^{-9}$~Hz (fig.~\ref{gw-spectrum-lep}). 
Observing such a deviation from the standard PGW
relic density at vanishing lepton asymmetry will elucidate and shed
new light on possible solutions for the existence of the baryon asymmetry of the
universe.
 
Finally, based on our current knowledge of the QCD phase diagram, the presence
of uncertainties in lattice simulations and our ignorance about the properties
of the quark gluon plasma in the early universe we do not have a unique and
confirmed picture about the real nature of QCD at early eras. The observation of
stochastic GWs produced by inflation may illuminate these issues
and deepen our understanding about the EoS of matter before BBN. 
The structure of the QCD phase can also be changed due to nonvanishing 
isospin chemical potential of charged pions or other SM
particles and may lead to pion condensation \cite{Abuki:2009hx,Brandt:2018bwq} which might affect the
thermal history of the universe and also the PGWs. We leave this
investigation for future work.


\begin{acknowledgments}

  We thank Dominik Schwarz, Dietrich B\"odeker for useful discussions and also Pasi
  Huovinen and Zoltan Fodor for providing the data of their equation of states. FH is grateful to Ken'ichi
  Saikawa for clarifications related to his work, ref.~\cite{Saikawa:2018rcs},
  and Nicolas Bernal for useful discussions. MMW acknowledges the support by 
  Studienstiftung des Deutschen Volkes. All authors acknowledge support
  by the Deutsche Forschungsgemeinschaft (DFG, German Research Foundation)
  through the CRC-TR 211 `Strong-interaction matter under extreme conditions'
  project number 315477589-TRR 211.
  
\end{acknowledgments}


\bibliography{qcd-gw.bib}

\end{document}